\documentclass[twocolumn,preprintnumbers,showpacs]{revtex4}
\usepackage{psfrag,graphicx,rotate,bm}
\begin{document}
\title{Oersted fields and current density profiles in spin-torque driven magnetization dynamics -- Finite element modelling of realistic geometries} 
\author{Riccardo Hertel}
\email{r.hertel@fz-juelich.de}
\affiliation{Institute of Solid State Research, IFF-9, J\"ulich Research Center, D-52425 J\"ulich, Germany}  
\newcommand{\be}{\begin{equation}}
\newcommand{\ee}{\end{equation}}
\newcommand{\ber}{\begin{eqnarray}}
\newcommand{\eer}{\end{eqnarray}}
\newcommand{\p}{\partial}
\newcommand{\heff}{\bm{H}_{\rm eff}}
\newcommand{\dmdt}{\frac{{\rm d}\bm{M}}{{\rm d}t}}
\newcommand{\djdt}{\frac{{\rm d}\bm{J}}{{\rm d}t}}
\begin{abstract}
The classical impact of electrical currents on magnetic nanostructures
is analyzed with numerical calculations of current-density
distributions and Oersted fields in typical contact geometries. 
For the Oersted field calculation, a hybrid finite element / boundary
element method (FEM/BEM) technique is presented which can be applied
to samples of arbitrary shape.
Based on the FEM/BEM analysis, it is argued that reliable
micromagnetic simulations on spin-tranfer-torque driven magnetization
processes should include precise calculations of the Oersted field,
particularly in the case of pillar contact geometries. Similarly,
finite-element simulations demonstrate that numerical
calculations of current-density distributions are required, e.g., in
the case of magnetic strips with an indentation. Such strips are
frequently used for the design of devices based on current-driven
domain-wall motion. A dramatic increase of the current density is
found at the apex of the notch, which is expected to strongly
affect the magnetization processes in such strips.     
\end{abstract}
\maketitle
\section{Introduction}

Current-driven magnetization processes have recently evolved to
probably the most active field in magnetism. Recent discoveries
demonstrating that, by exploiting the electron spin, the magnetization
state of a nanomagnet could be influenced directly by electrical
currents instead of external magnetic fields
\cite{Kiselev03,Rippard04,Tsoi03,Vernier04} have given rise to
numerous extensive studies on the spin-transfer torque (STT) effect.
Magnetization dynamics induced by STT \cite{Slonc96, Berger96} is
attractive for both technological aspects and fundamental
physics. Unlike the field-induced magnetic switching, which in the
case of nanodevices is connected with the technological difficulty
of focussing  magnetic fields on very small length scales, the
STT effect provides the possibility to switch, e.g., individual
magnetic nanoelements in integrated circuits or in  densely packed 
arrays \cite{Liu07}. Moreover, the current-induced magnetization
dynamics differs   
qualitatively from the field-induced dynamics. A striking example
thereof is the excitation of high-frequency 
magnetic oscillations 
induced by means of DC
spin-polarized currents \cite{Kiselev03,Rippard04}; an effect without
analogy in the field-driven dynamics. The further possibility of
tuning the frequency  of these oscillations by varying the current
strength bears a high potential for future applications.

The STT effect and the related fascinating features of
current-induced magnetization dynamics have been extensively studied
by various groups over the last years. Micromagnetic simulations have 
nowadays developed to such a high level of accuracy and reliability
that they provide an important tool for the design and for the
understanding of magnetic nanostructures and their magnetic
properties. While the STT effect has been implemented in several
micromagnetic codes \cite{Berkov06,Lopez05,LeeDeac04}, there is often still a
quantitative discrepancy between  simulated and experimental results
in the case of current-induced magnetization dynamics
\cite{BerkovMiltat07}. This differs from the static \cite{Cherifi05,
  Hertel05} and the field-driven cases \cite{Buess04}, where countless
examples can be found showing excellent agreement between experiment
and simulation, thereby demonstrating the predictive power of
classical micromagnetic simulations.  

One of the possible sources for such disagreement 
is connected 
with the Oersted field. Current-induced magnetization processes 
require high current densities, typically larger than about
10$^{11}$ A/m$^2$. Such current densities are inevitably connected
with sizable magnetic fields, known as Oersted or Amp{\`e}re fields
\cite{Jackson99}. In addition to the usual external and internal
effective fields entering the Landau-Lifshitz-Gilbert equation of
motion, and in addition to the STT term, these Oersted field also
affect the magnetization. The importance of the Oersted field in
spin-transfer induced magnetic switching process has recently been
pointed out by various groups \cite{Stoehr06,Devolder07}.
In view of the aforementioned high accuracy of classical micromagnetic
simulations, a numerical method is required in order to 
calculate the Oersted field contribution with high precision when STT
effects are studied. In this article a method based on a combination
of the finite-element method (FEM) and the boundary element method
(BEM) is presented which allows for fast and precise calculations of
the Oersted field in arbitrary contact geometries. The hybrid FEM/BEM 
scheme has several advantageous features, including the simple
consideration of inter-particle interactions resulting from 
Oersted fields generated, {\em e.g.}, in complex nanocircuits
\cite{Parkin04} or arrays of current-carrying nanopillars
\cite{Kaka05}.   

The paper is structured as follows. First, the basic equations and the
numerical method for the Oersted field calculation are
peresented. Subsequently, in section \ref{jdense} a 
method to calculate current-density distributions in non-trivial
geometries is outlined. Finally, a few examples on the application of
these methods 
to structures that are currently used for intensive studies on 
current-driven magnetization processes are presented in section
\ref{examples}. 
\section{Basic equations}
Consider a current-carrying sample of known size and shape. 
Let us further assume that the current density distribution
$\bm{j}(\bm{r})$ is known for every point $\bm{r}$ in the volume
$V$ of sample. The calculation of $\bm{j}(\bm{r})$ will be discussed in 
in section \ref{jdense}.
The Oersted field $\bm{H}_c$ connected with a current density
distribution $\bm{j}(\bm{r})$ can be obtained by direct integration over the volume of the current-carrying particle: 
\ber\label{direct}
\bm{H}(\bm{r})&=&\bm{\nabla}\times\frac{1}{4\pi}\int\frac{\bm{j}(\bm{r}')}
{\left|\bm{r}-\bm{r}'\right|}\,{\rm d}V'\\
&=&\int\bm{j}(\bm{r}')\times\left(\bm{\nabla}'G\right)\,{\rm d}V'
\eer
where
\be
G(\bm{r}-\bm{r}')=\frac{1}{4\pi}\frac{1}{\left|\bm{r}-\bm{r}'\right|}
\ee
is Green's function with
\be\label{deltagreen}
\Delta G=-\delta(\bm{r}-\bm{r}')
\ee

and $\delta(\bm{r}-\bm{r}')$ is the Dirac delta function. Note that
$\bm{\nabla}'G=-\bm{\nabla}G$. 
If the volume of the sample is discretized into finite elements, it is
relatively simple to perform such a numerical integration. However,
the direct integration according to eq.~(\ref{direct}) has a number of
disadvantages from a numerical point of view. The volume integration
over Green's function may give rise to serious problems of accuracy: 
Simple integration schemes  can lead to inaccurate results because the
value of Green's function can vary significantly within a single
finite element containing a point $\bm{r'}$ if the viewpoint $\bm{r}$
is close to that element. Moreover, the computational effort involved
with this direct integration is very large: The calculation of each
component of  
$\bm{H}_{\rm c}$ at a single point $\bm{r}$ requires an integration over
the whole volume $V'$ of the conductor. A more convenient approach
in terms of accuracy and efficiency consists in 
numerically solving 
partial differential equations instead of the direct integration
according to eq.~(\ref{direct}). In the following, the basic steps
for a  FEM/BEM formulation of the Oersted field problem are
outlined.  

The fundamental equation for the calculation of Oersted fields is
Amp{\`e}re's law:
\be
\bm{\nabla}\times\bm{H}=\bm{j}
\ee
In a current-carrying ferromagnet with magnetization $\bm{M}$, the
magnetostatic field $\bm{H}$ is constituted by two parts:  
\be
\bm{H}=\bm{H}_{\rm s}+\bm{H}_{\rm c}
\ee
where $\bm{H}_{\rm s}$ is the demagnetizing field (also called dipolar
field of stray field) created by the magnetic
surface and volume charges, and the field $\bm{H}_{\rm c}$ is the
magnetic Oersted field (also called Amp\`ere field) due to the current density
$\bm{j}$.  The following equations apply to these two static magnetic
field contributions: 
\ber
\bm{\nabla}\times\bm{H}_{\rm s}=\bm{0}\hspace{1cm}&&\hspace{1cm}
\bm{\nabla}\times\bm{H}_{\rm c}=\bm{j}\label{curlH}\\
\bm{\nabla}\bm{H}_{\rm s}=-\bm{\nabla M}
\hspace{1cm}&&\hspace{1cm}
\bm{\nabla H}_{\rm c}=0\label{divH}
\eer
The calculation of $\bm{H}_{\rm s}$  is one of the central parts of
any micromagnetic code. Several powerful methods have been discussed
for the numerical calculation of the dipolar field
\cite{Chen97,Yuan92} and this part of the problem can now be generally
considered as solved. 
Due to its
similarity with the method described later,  
one particular efficient method to calculate
the demagnetizing field should be pointed out here, namely the hybrid
FEM/BEM algorithm presented by Koehler and
Fredkin~\cite{Koehler90}. Henceforth, only the contribution from the  
electric  current $\bm{H}_{\rm c}$ shall be considerd, and the
subscript ``c'' is omitted for simplicity.   

Equations (\ref{curlH}) and (\ref{divH}) yield
\be\label{Poisson}
\Delta\bm{H}= -\bm{\nabla}\times\bm{j}
\ee
In the region outside the conductor, the source term is zero, i.e.,
\be\label{Laplace}
\Delta\bm{H}= \bm{0}
\ee
Thus, each component of $\bm{H}$ satisfies an equation of the Poisson
form inside the conductor and of the Laplace form on the
outside. Equations (\ref{Poisson}) and  (\ref{Laplace}) describe an
{\em open boundary problem}. Such problems are characterized by the
absence of well-defined boundary conditions at the sample surface. The
solution is uniquely defined by the condition of ``regularity at
infinity'', {\em i.e.}, a Dirichlet-type of boundary condition
for $r\to\infty$:  
\be
\lim_{r\to\infty}\bm{H}(\bm{r})=\bm{0}
\ee

A possible approach for the consideration of such boundary
conditions consists in attempting to expand the computational region
to ``infinity'', {\em e.g.} by applying bijective transformations to
map the infinite volume surrounding the sample onto a volume of finite
size, which can then be discretized with finite elements \cite{Brunotte92}.
Such transformation methods can suffer from accuracy problems 
because the discretized spatial transform effectively corresponds to a
truncation of the computational region at more or less large
distances. Moreover, these methods require the external region to be
free of charges, thereby precluding the possibility of calculating
the field in the case of interacting current carriers. An alternative
approach, which solves these problems, is the use of a hybrid finite
element /  boundary element method. In the boundary element
method, the fundamental solution of a partial differential equation is 
used. In this case, the fundamental solution is given by Green's
function, which automatically fulfils the boundary condition at
infinity. 

\section{Hybrid FEM/BEM formulation\label{formula}}
The first step in developing the FEM/BEM formulation consists in an
analysis of the properties of the solution at the sample boundary. 
From eq.~(\ref{Poisson}) a jump condition for the normal derivative of
$\bm{H}_{\rm c}$ at the boundary of the current carrier is easy to derive:
\be\label{jump}
\bm{n}\left(\left.\bm{\nabla}H_x\right|_{\rm in}-
\left.\bm{\nabla}H_x\right|_{\rm
  out}\right)=\bm{n}\left(\bm{e}_x\times\bm{j}\right)
\ee
This equation correspondingly holds also for the $y$ and $z$
components of $\bm{H}$. In eq.~(\ref{jump}) $\bm{e}_x$ is the unit
vector along the $x$ axis and $\bm{n}$ is the surface unit vector
directed outwards. Note that this condition represents a discontinuity
of the gradient of $H_x$ at the boundary. However, it does not provide
any information  on the value of the inward limit of the gradient of
$H_x$. This condition can therefore not be used directly as a Neumann
boundary condition at the boundary $\p\Omega$ as it would be required
for a unique solution of the differential equation in the region 
$\Omega$.   

In order to obtain useful boundary conditions at the sample surface,
the Oersted field can be split in two parts, 
$\bm{H}=\bm{H}^{(1)}+\bm{H}^{(2)}$. These fields shall have the following
properties:
(i): Outside the current carrier, the part $\bm{H}^{(1)}$ is zero. 
(ii): The part $\bm{H}^{(1)}$ satisfies Poisson's equation 
  \be\label{part1pos} 
\Delta H^{(1)}_x= -\bm{e}_x\cdot\left(\bm{\nabla}\times\bm{j}\right)
\ee
with Neumann
  boundary conditions
\be\label{part1jump}
\bm{n}\left(\left.\bm{\nabla}H^{(1)}_x\right|_{\rm in}
\right)
=\bm{n}\left(\bm{e}_x\times\bm{j}\right)
\ee
(iii): The part $\bm{H}^{(2)}$ satisfies the Laplace equation
\be\label{part2lap}
\Delta H^{(2)}_x= 0
\ee
and its derivatives are continuous along the boundary.

The boundary conditions required for the solution of
eq.~(\ref{part2lap}) will be discussed later. 
The jump condition for the individual components $H^{(1)}$ and
$H^{(2)}$ at the boundary
\be
\left.H^{(1)}_x\right|_{\rm in}+\left.H^{(2)}_x\right|_{\rm
  in}=\left.H^{(2)}_x\right|_{\rm out}
\ee
results from the condition that the field $\bm{H}$ must be continuos
at the surface.
\begin{figure}
\centerline{
\includegraphics[width=\linewidth]{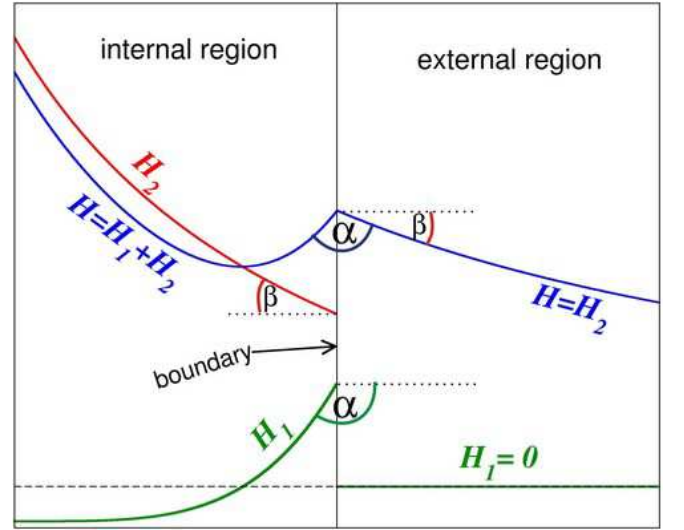}
}
\caption{\label{split}(Color online) The field $\bm{H}$ is split in two parts. The
  angle $\alpha$ represents the known value of the discontinuity of
  the derivative of $\bm{H}$ at the boundary (eq.~\ref{jump}). The
  part $H_1$ of the solution is chosen such that its gradient at the 
  boundary corresponds to $\alpha$. Moreover, $H_1$ is equal to
  zero in the external region. Both, $H_1$ and $H_2$ are generally
  discontinuous at the boundary. The sum, i.e., the field $H$ is
  however continuous. The gradient of the part $H_2$ is constant along
  the boundary, as sketched by the angle $\beta$.} 
\end{figure}
The splitting of the field $H_x$ in two parts is schematically shown
in  Fig.~\ref{split}. The advantage of this splitting is 
that it is possible to extract  
useful boundary conditions required for the
solution of Poisson's equation: By setting 
$H^{(1)}_x$ to zero outside the sample, the discontinuity
condition (\ref{jump}) is converted into a Neumann boundary   
condition (\ref{part1jump}) that can be used to solve
eq.~(\ref{part1pos}). With the finite element method, it is relatively
easy to solve the Poisson equation (\ref{part1pos}) with given Neumann 
boundary conditions (\ref{part1jump}). After that, the somewhat
more complicate part needs to be addressed, {\em i.e.}, finding the 
values of $\bm{H}^{(2)}$ at the particle surface. These values will
provide the Dirichlet boundary conditions required for the 
solution of Laplace's equation (\ref{part2lap}). For this task, the
boundary element method is used.  

Multiplying eq.~(\ref{deltagreen}) with $\bm{H}_1$ and eq.~(\ref{part1pos}) with $G$ 
yields, after subtraction  and integration over the volume $V$ of the region $\Omega$:  
\ber
\int G\Delta H_1^x\,{\rm d}V'&-&\int H_1^x\Delta G\,{\rm d}V'\nonumber\\&=&
\int H_1^x\delta(\bm{r}-\bm{r'})\,{\rm d}V'
-\int\bm{e}_x\left(\bm{\nabla}'\times\bm{j}\right)G
\,{\rm d}V'\nonumber\\
\eer
which, by virtue of Green's theorem, transforms into 
\ber
\oint G\frac{\p H_1^x}{\p\bm{n}}\,{\rm d}S'&-&\oint H_1^x\frac{\p
  G}{\p\bm{n}}\,{\rm d}S'=
\int H_1^x\delta(\bm{r}-\bm{r}')\,{\rm d}V'
\nonumber\\
&+&\oint\bm{n}\left(\bm{e}_x\times\bm{j}\right)G\,{\rm d}S'-\bm{e}_x
\int\bm{j}\times\left(\bm{\nabla}'G\right)\,{\rm d}V'\nonumber
\eer
The surface integrals $\oint\,{\rm d}S'$ extend over the boundary $\p\Omega$ of the region $\Omega$, {\em i.e.}, the sample surface. 
Inserting the boundary condition for $H_1$
\be\label{jump3}
\frac{\p H_1^x}{\p\bm{n}}=\bm{n}\left(\bm{e}_x\times\bm{j}\right)
\ee
the equation simplifies to 
\be
-\oint H_1^x\frac{\p G}{\p\bm{n}}\,{\rm d}S'
=\int H_1^x\delta(\bm{r}-\bm{r}')\,{\rm d}V'
-\bm{e}_x
\int\bm{j}\times\left(\bm{\nabla}'G\right)\,{\rm d}V'
\ee
The last term can be identified as the right-hand side of eq.~(\ref{direct}), 
yielding 
\be\label{almostdone}
-\oint H_1^x\frac{\p G}{\p\bm{n}}\,{\rm d}S'
=\int H_1^x\delta(\bm{r}-\bm{r}')\,{\rm d}V'
-\left[H_1^x(\bm{r})+H_2^x(\bm{r})\right]
\ee
The integral $\int H_1^x\delta(\bm{r}-\bm{r}')\,{\rm d}V'$ is trivial
as long as the point $\bm{r}$ is located inside or outside the volume
$V$. The situation requires more attention when $\bm{r}$ is a point on
the boundary. In this case, the inward limit is taken by introducing
an infinitesimal distance of the point (located in the internal
region) to the surface. The resulting integral over the delta  
function is in this case
\be
\int H_1^x\delta(\bm{r}-\bm{r}')\,{\rm d}V'=\frac{\Psi}{4\pi}H_1^x(\bm{r})
\ee
where $\Psi$ is the solid angle subtended at the boundary point
$\bm{r}$. 

Hence, an equation is obtained with which $H_2^x(\bm{r})$ can be
calculated at each boundary point:
\be\label{finalBE}
H_2^x(\bm{r})=\oint H_1^x(\bm{r}')\frac{\p G}{\p\bm{n}}\,{\rm
  d}S'+\left(\frac{\Psi}{4\pi}-1\right) H_1^x(\bm{r})
\ee
Accurate numerical methods to perform this integral by means of BEM
are discussed in Ref.~\cite{Lindholm84}.  
In principle, eq.~(\ref{almostdone}) would be sufficient to calculate
$\bm{H}_2$ at any point inside (and outside) the volume, so that the
resulting total field $\bm{H}=\bm{H}_1+\bm{H}_2$ could be
calculated at any discretization point. However, the calculation 
of $\bm{H}_2$ at one point $\bm{r}$ requires in this case an
integration over the whole surface.  This approach would thus
have similar disadvantages as the direct integration according to
eq.~(\ref{direct}). By using equation (\ref{finalBE}) instead, we
obtain at relatively low cost the boundary values of
$\bm{H}_2$. Having these Dirichlet boundary conditions, it is easy 
to solve eq.~(\ref{part2lap}) in the volume $V$ by means of the FEM.

It is noteworthy that eq.~(\ref{finalBE}) has the same form as it has
been used in the FEM/BEM scheme described in Ref.~\cite{Koehler90} to
calculate the scalar magnetic potential of a ferromagnet. From the
viewpoint of implementation into a program, this means that the 
matrix required for the numerical calculation the values of $\bm{H}_2$
according to eq.~(\ref{finalBE}) is already available in the
micromagnetic code if the FEM/BEM scheme presented in
Ref.~\cite{Koehler90} is used for the calculation of the magnetic scalar
potential. 
\section{Calculation of current density distributions\label{jdense}}
Unless the geometry of the current-carrying sample is trivial, the
current density distribution $\bm{j}(\bm{r})$ needs to be determined
numerically prior to the calculation of 
the Oersted field. The starting point for the calculation of the
current density distribution is Ohm's law
\be
\bm{j}=\sigma\bm{E}
\ee
$\bm{E}$ is the local electric field and the conductivity $\sigma$ is
assumed to be a scalar. The electric field is the  
gradient field of the electrostatic potential $U$, so that
$\bm{E}=-\bm{\nabla}U$. 
Charge conservation yields $\bm{\nabla}\bm{E}=0$ and thus $
\bm{\nabla  j}=0$, which ultimately leads to 
\be\label{statdiff}
\bm{\nabla}\left(\sigma\bm{\nabla}U\right)=0
\ee
This equation has the form of the stationary  diffusion
equation and converts into the Laplace equation for $U$ if $\sigma$ is
homogeneous. 

Elliptic differential equations of the type (\ref{statdiff}) are
routinely solved numerically with FEM. 
 However, appropriate boundary
conditions must be specified to obtain a unique solution. 
In this case, the boundary conditions are given by the fact that the
current is not flowing perpendicular to the sample surface
$\bm{j}\cdot\bm{n}=0$
except for the leads, where a known current density is entering and
leaving the sample. In the case of two contact leads of the same size,
the boundary condition is $\bm{j}\cdot\bm{n}=\pm j_0$ at the
leads, leading to
\be
\frac{\p U}{\p n}=\left\{\begin{array}{lcl}
\mp j_0/\sigma&&{\rm at\,\,the\,\,leads}\\
&&\\
0&& {\rm at\,\,the\,\,rest\,\,of\,\,the\,\,boundary}
\end{array}
\right.
\ee
In a more general case, only the value of the total current $I$ flowing
through the sample is known. The value of the current density at
the positive and negative leads then needs to be determined by
dividing the total current flowing through each lead by the area of
the contact region. 
Note that eq.~(\ref{statdiff}) with given Neumann boundary conditions 
is as easy to solve numerically for homogeneous conductivity
$\sigma$ as it is in the case of inhomogeneous conductivity
$\sigma(\bm{r})$.  

The method described here for the calculation of current density
distributions is generally well known and has been applied in various
previous publications \cite{Moussa01,Holz03,Sarau07}. It is 
included in this manuscript mainly for completeness, because such
current density distribution calculations usually are a necessary
prerequisite  for the calculation of Oersted fields.  
\section{Examples\label{examples}}
The methods outlined above can be applied to various
problems which are of high importance for modern research  topics in
the field of nanomagnetism. In this section, a few examples are
presented. These include the current density distribution in a thin
strip with an indentation (``notch'') and the calculation of the 
current density distribution and of the Oersted field in a pillar contact
geometry as it is used to study high-frequency excitations in
nanomagnets \cite{Kiselev03,Dassow06}.
\subsection{Current density distribution in a thin strip with a notch}

Ferromagnetic thin strips with width in the sub-$\mu$m range and
thicknesses of the order of a few ten nm have attracted much interest over 
the past years. One reason is the particular type of head-to-head domain
walls that occur in such structures \cite{McMichael97b}. Magnetic
strips have been proposed to be used in future nanomagnetic devices,
in which the head-to-head domain walls would serve as units of
information that could be processed in logical devices
\cite{Parkin08,Allwood02}. It has further been shown that the domain walls in
such strips can be displaced by means of the STT effect if an electric
current of sufficient strength is flowing through the sample
\cite{Tsoi03,Vernier04}. Since the domain walls can be displaced
continuously along a magnetic strip, it is of practical importance to
gain control of their position, so that they are located and displaced
between well-defined positions in the strips. 
This can be achieved by means of indentations or notches in the
strips, which act as an attractive potential for the domain
walls \cite{Klaui04b}. Depending on their type, the domain 
walls are then either located exactly at the notch or in its close
vicinity \cite{Parkin08b,Klaui04b}. 

The influence of notches on the
micromagnetic configuration has been studied intensively and reported
in several publications. An equally important question for the study
of current-driven domain wall motion in such indented strips is the
influence of a notch on the current density distribution.

\begin{figure}
\centerline{
\includegraphics[width=\linewidth]{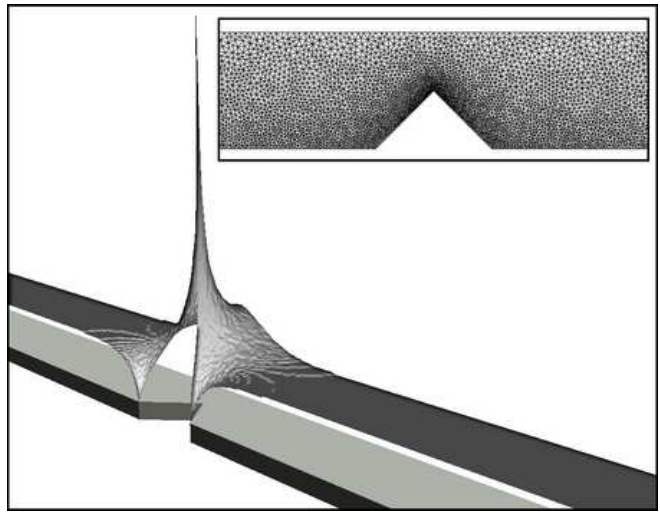}
}
\caption{\label{jnotch}Current density distribution in a thin, flat 
  strip with a notch. The inset shows a part of the finite-element 
  mesh used for the analysis close to the constriction. The main image 
  shows a perspective view  on the region of interest of the indented 
  strip and a topographical representation of the magnitude of the
  local current density. The strip is contacted such that at large
  distances from the notch the current flows parallel to the
  strip. The local current density is dramatically increased near the
  apex of the notch. Contratry to this, the current density on the
  flat side of the constriction is only slightly elevated compared
  with the value in the unconstrained parts of the strip.}      
\end{figure}

To address this question, the  current
density distribution can be simulated with finite element modelling
according to the method described in the previous section.   
The model used for the simulation is a 1.2\,$\mu$m long strip (width:
100\,nm, thickness: 20\,nm) with an indentation in the center. The
indentation reduces the width of the strip to 50\,nm in the narrowest
part. The angle of the indentation is 45$^\circ$. The region of
interest of the finite element mesh is shown in the inset of
Fig.~\ref{jnotch}. In the vicinity of the notch, the tetrahedral mesh
is locally refined to increase the numerical accuracy. A total current
of  $I=2$\,mA is flowing along the wire, and a homogeneous electric
conductivity is assumed. 

The topographical representation in
Fig.~\ref{jnotch} displays the local value of the current density. As
expected, the current density increases at the constriction. Much more
significant than the reduction of the width is, however, the 
impact of the notch. Where the wire is not constrained, the current
density is $j=10^{12}$\,A/m$^2$. At the apex of the notch, the value
increases drastically to $j=5.5\cdot 10^{12}$\,A/m$^2$, while in the
opposite, flat part of the strip it only increases up to $j=1.55\cdot
10^{12}$\,A/m$^2$. Hence, in the narrow part of the strip, the current 
density distribution is highly inhomogeneous. Over the small distance
of 50\,nm it changes almost by a factor of four.  The values
of the local current density obviously scale linearly with the applied
current. The value of 2\,mA has only been chosen here as an example
since the resulting current density values are of the order of those
reported in corresponding experimental studies. The current density
{\em profile} is independent of the value of the applied
current. Hence, the inhomogeneities of the current density
distribution are directly connected with the sample geometry, and not
with the value of the applied current. The profile is moreover
invariant with respect to scaling.

The knowledge of this drastically inhomogeneous current density
distribution at a notch is expected to be of utmost importance for the
design and for the understanding of devices based on current-driven
domain-wall displacement. The huge differences of the local current
density in the constriction region are expected to have a significant
impact on the local strength of the STT effect, the onset of
electromigration processes and on the heating of the sample. The
seemingly plausible  approximation of homogeneous current density
distribution in the constriction region hence clearly appears to be inadequate, 
and should be dropped.   

\subsection{Oersted field calculation - comparison with analytics}
The method presented in section \ref{formula} to calculate the Oersted
field for a given current density distribution can be tested by
comparing the numerical results with known analytic solutions.
A simple example is the
magnetic field of an infinitely long current-carrying cylinder with
homogeneous current density. The current is flowing parallel to the
symmetry  axis. If a wire with radius $R$ is parallel to the $z$
direction, with the current $I$ flowing along $\bm{e}_z$ and $\phi$
and $\rho$ being the  azimuthal and radial coordinate, respectively,
the Oersted field is 
\be
\bm{B}(\bm{r})=\mu_0H(\rho)\bm{e}_\phi=\frac{\mu_0I}{2\pi}\bm{e}_{\phi}\left\{
\begin{array}{lcr}
\rho / R^2 & & (\rho \leq R)\\&&\\
1/\rho & & (\rho>R)
\end{array}
\right.
\ee
\begin{figure}
\centerline{
\includegraphics[width=\linewidth]{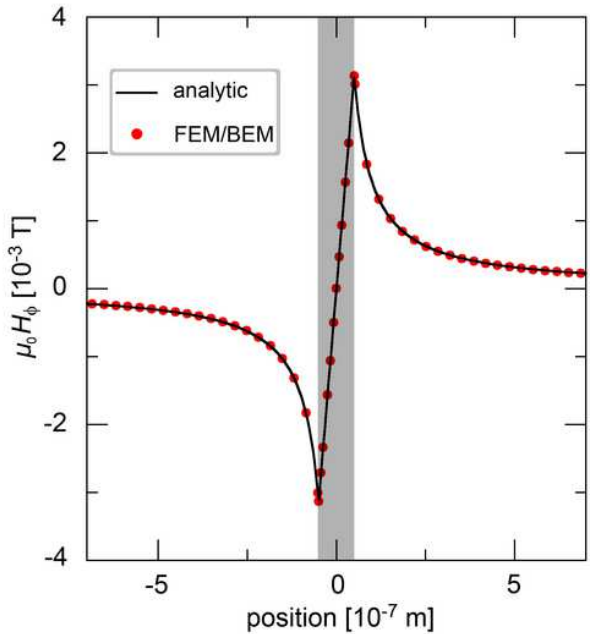}
}
\caption{\label{compare}(Color online) Oersted field of a long, current-carrying
  cylinder with  homogeneous current density. The current is flowing
  parallel to the wire axis. In this case, the field $\bm{H}$ only
  has an azimuthal component. The computed values (dots) perfectly
  match with the analytic solution  (line). The grey area denotes the 
  region inside the wire. }  
\end{figure}
In Fig. \ref{compare} this analytical result (solid line) is compared
with the computed values (dots) 
resulting from the FEM/BEM simulation in the case of an 8$\mu$m long wire
with $R=50$\,nm and homogeneous current densitiy
$\bm{j}=10^{11}$A/m$^2\cdot\bm{e}_z$. Excellent agreement is
obtained. Numerical tests show that minor deviations can be
further reduced by increasing the discretization density, but are not
of practical importance. This example proves the correctness and 
accuracy of the hybrid FEM/BEM scheme. Obviously, the advantage of the
FEM/BEM algorithm is its applicability to samples of complex shape,
which cannot be calculated analytically. Such an example is given
in the following subsection, where the current density distribution and
the resulting Oersted field are calculated for a complex contact
geometry as it is used to study current-induced stationary
high-frequency excitations of nanomagnets. 

\subsection{Current densities and Oersted fields in a nanopillar contact
  geometry} 
A remarkable difference between the magnetization dynamics induced by  
spin-polarized electrons as compared to the field induced dynamics is
the possibility of generating high-frequency, stationary oscillations
in a nanomagnet with a DC current. This effect, which can occur when a
sub-micron sized ferromagnetic thin-film element is exposed to a
spin-polarized current flowing perpendicularly through its surface,
has been predicted \cite{Slonc96,Berger96} theoretically and confirmed
experimentally \cite{Kiselev03}. In experimental setups, the thin-film
element is usually embedded into a pillar-shaped multilayer
nanostructure, which is contacted by mesoscopic leads on the top and
on the bottom, so that the electric current flows parallel to the pillar
axis. Numerous experimental and numerical studies on the
current-induced magnetization dynamics in nanomagnets within a pillar
contact have been reported recently (see, e.g., 
\cite{MiltatStiles06} and references therein).   
While the STT effect is driving the magnetization dynamics, it 
should be kept in mind that the Oersted field connected with the
electric current may represent a non-negligible
perturbation which could affect the magnetization
dynamics. Considering that also in this case the typical current 
densities are of the order of $10^{12}$ A/m$^2$, the Oersted field is
expected to provide a sizable contribution even though the pillar
dimensions are in the sub-micron range.
\begin{figure}
\centerline{
\includegraphics[width=\linewidth]{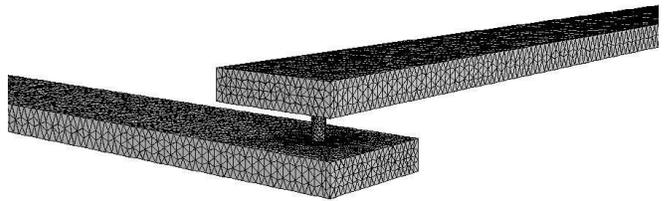}
}
\caption{\label{themesh}Perspective view on the finite-element mesh used for the
  contact geometry. The ends of the 4$\mu$m long contacting strips,
  where the current is flowing into and out of the sample, are located
  outside of this frame, which is a magnified view on the region of
  interest where the pillar is located. The mesh is locally refined 
  at the nanopillar.}
\end{figure}

\begin{figure}
\centerline{
\includegraphics[width=\linewidth]{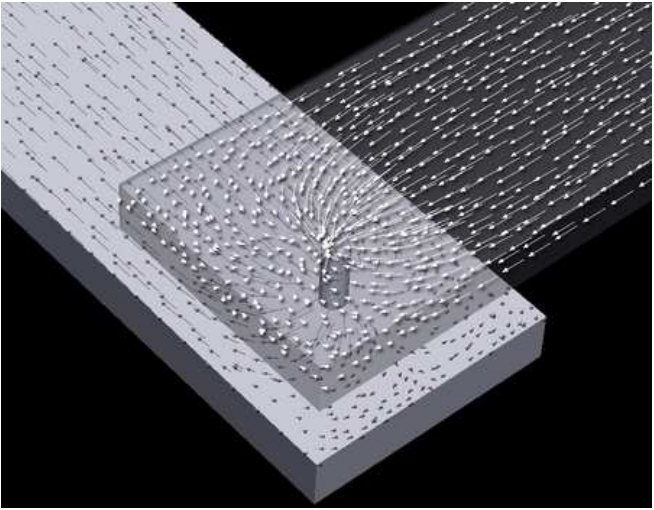}
}
\caption{\label{jleads}Simulated current density distribution in the
  contact geometry. The arrows represent the direction of the local
  current, and their length is proportional to the local value of the
  current density. For better visibility, only a selection of the
  computed data is shown. For the same reason, the current density
  distribution inside the pillar is not shown, since the much larger
  current density there would result in very long arrows in this
  representation. The top electrode is displayed as semi-transparent
  in order to visualize the current density distribution inside the
  leads.    }      
\end{figure}
To obtain precise values of the Oersted field in such complicated
contact geometries, 
the approach consists in first
simulating the current density distribution in the contact and
subsequently calculating the magnetic field. Apart from the current
flowing through the pillar, also the current in the leads 
(which act as what has become known as ``strip lines'' in the case of
field-induced magnetization dynamics) contributes to the total
magnetic field acting on the nanomagnet. Fig.~\ref{themesh} shows the 
finite-element mesh used to simulate the current and field
distribution in a pillar contact geometry. In this example, the pillar
diameter is 50\,nm and the pillar height is 120\,nm. It is contacted
by  strips of 500\,nm width and 100\,nm thickness. Each contacting
strip is 4\,$\mu$m long and the top and bottom electrodes form an
angle of 90$^\circ$. The geometry roughly corresponds to typical 
experimental setups \cite{Dassow06}, even though the contacting strips
are often wider than shown here. The purpose of this
simulation is rather to provide an example for the possibility of
calculating current density distributions and Oersted fields in
realistic geometries than to reproduce all details of a specific
experimental setup.

A current of 2\,mA is flowing through the sample along the
contacting strips. 
Since the total value of the current and the diameter of the pillar are
known, the current density inside the pillar can be trivially calculated
without any simulation (yielding $j\simeq 10^{12}$\, A/m$^2$). The
same holds for the current density in the 
contacting strips at regions sufficiently far away from the
contact ($j=4\times 10^{10}$\, A/m$^2$). The more interesting
situation occurs in the leads in the vicinity of the contact to the
pillar. The complex current density distribution in this
region is shown in Fig.~\ref{jleads}. 

\begin{figure}
\centerline{
\includegraphics[width=\linewidth]{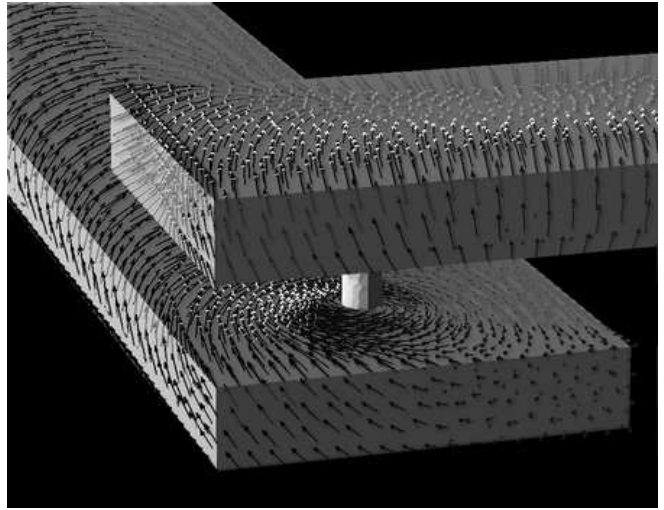}
}
\caption{\label{oleads}Computed Oersted field resulting from the
  current distribution shown in Fig.~\ref{jleads}. The circulation of
  the magnetic field around the current-carrying regions is clearly
  visible. The black arrows have a component directed outside of the
  sample, while the pale gray arrows are pointing towards the
  inside.  Only the magnetic field at the surfaces of the leads is
  displayed. The magnetic field of the pillar is not shown for better
  visibility of the overall structure of the field.}     
\end{figure}

The discretized form of the computed three-dimensional vector field
$\bm{j}(\bm{r})$ can now be used as an input for the Oersted field
calculation. By applying the FEM/BEM algorithm we then obtain the 
Oersted field in the whole contact geometry. 
The result is displayed in Fig.~\ref{oleads}, where the field
circulation around the contacting strips and, in particular, around
the pillar can clearly be seen. 
For clarity, only the field in the contacting
strips is displayed. 

For such experiments and their
correct interpretation, the field distribution inside the pillar is
decisive. 
The profile of the field inside the pillar is very similar to the one
displayed in Fig.~\ref{compare}: The dominant component of the field
is azimuthal, with a magnitude that inside the pillar increases linearly
with the distance from the central axis. The  peak value at the
boundary of the pillar is in this case 28\,mT, which certainly
represents a non-negligible field. Compared with the field strength due
to the current flowing in the pillar, the effect due to the current
in the contacting strips can be considered as a perturbation,
which provides only an asymmetry along a direction at 45$^\circ$ with  
respect to the strips. Due to this asymmetry, the magnitude of the
maximum and the minimum value of both, the $x$ and the $y$ component
of the Oersted field differ by about 12\%. 
Whether the influence of the current through the leads can be
neglected depends on the specific experimental setup and the problem
that is being investigated. 
While the field due to the contacting strips may in several cases be
safely neglected, it is of essential importance to consider the
precise {\em height} of the pillar, since it has a direct and decisive
impact on the strength of the Oersted field. 
Reliable calculations of Oersted fields in nanopillar geometries can
only be obtained if the height of the current-carrying pillar
is known. More details on the influence of the
pillar height on the Oersted field and its impact on current-driven
magnetization dynamics will be discussed in a forthcoming article. 
\section{conclusion}
The classical interaction of electric currents with the magnetization
of ferromagnets is given by the Oersted field. Its importance should
not be underestimated in theoretical studies on the emerging field of 
current-induced dynamic magnetization processes driven by the STT
effect. As shown in this study, the values of the Oersted field
obtained by using finite-element models of typical experimental
contact geometries can be as high as a few tens of mT. In view of
these large fields, it appears likely  that the details of several
current-induced dynamic magnetization processes that are currently
studied by many groups are the result of a combination of both, the
STT effect and the Oersted field. Thus, for a reliable interpretation
of experimental data by means of micromagnetic simulations, the
Oersted field may {\em a    priori} not be neglected. The hybrid
FEM/BEM method presented in this article  provides a flexible and
accurate tool to calculate the Oersted field, and thus to clarify its
importance and its influence on the dynamics of nanomagnets
in STT studies. The value of the Oersted field depends sensitively on
the geometry of the contact setup. For instance, in pillar geometries,
the pillar height is of  decisive importance.

Also in the case of current-induced domain wall displacement in
ferromagnetic strips, some seemingly plausible simplifying assumptions
may not be applicable. Micromagnetic simulations should in this case
be extended by a routine to calculate the current-density distribution  
like, {\em e.g.}, the FEM scheme described in this article. On the
example of the frequently used case of indented thin strips, the
simulations have shown that geometric variations can have a dramatic
impact on the current-density distribution in thin strips. Such a
notch can give rise to a strongly inhomogeneous current-density
profile along the strip width  connected with a strong, localized
increase of the current density. It is therefore not sufficient to
assume that the increase of  current density in such an indented strip
simply correlates with the reduction of the cross-section at the notch.

\section*{Acknowledgements}
I am indebted to Prof.~C.M.~Schneider for valuable comments on the
manuscript and to Sebastian Gliga for precious help on the graphical
representation of the computed field and current distributions.

\bibliographystyle{phreport}

\end{document}